\documentclass[10pt,conference]{IEEEtran}
\makeatletter
\def\ps@headings{%
\def\@oddhead{\mbox{}\scriptsize\rightmark \hfil \thepage}%
\def\@evenhead{\scriptsize\thepage \hfil \leftmark\mbox{}}%
\def\@oddfoot{}%
\def\@evenfoot{}}
\makeatother
\usepackage{amssymb,color,amsmath,clock}
\usepackage{paralist}
\usepackage{graphicx,epsfig}
\usepackage{multicol}

\usepackage{arydshln}
\usepackage{graphicx}
\usepackage[urlcolor=rltblue,colorlinks=true]{hyperref} %
\definecolor{rltblue}{rgb}{0,0,0.75}
\usepackage{cite,manfnt}

\newcommand{\F}{\mathbf{F}}

\newcommand{\N}{\mathcal{N}}

\newtheorem{theorem}{\textbf{Theorem}}
\newtheorem{lemma}[theorem]{\textbf{Lemma}}
\newtheorem{definition}[theorem]{\textbf{Definition}}

\newtheorem{example}[theorem]{Example}

\newcommand{\nix}[1]{}

\begin{document}
\title{S-MATE: Secure Coding-based Multipath Adaptive Traffic Engineering}
\author{Salah A. Aly$^\dag$$^\ddag$~~~~~~ Nirwan Ansari$^\dag$~~~~~~Anwar I. Walid$^\S$~~~~~~H. Vincent Poor$^\ddag$ \\
 $^\dag$Dept. of Electrical \& Computer Engineering, New Jersey Inst. of Tech., Newark, NJ 07102, USA \\
$^\ddag$ Dept. of Electrical ~Engineering,~ Princeton University, ~Princeton, ~NJ 08544, ~USA\\
$^\S$Bell Laboratories~ \& Alcatel-Lucent, ~Murry Hill,  ~NJ 07974, ~USA}

 \maketitle

\begin{abstract}
There have been several approaches to provisioning  traffic  between core network  nodes  in Internet Service Provider (ISP) networks. Such approaches aim to minimize network delay, increase network capacity, and enhance network security services.  MATE (Multipath Adaptive Traffic Engineering) protocol has been proposed for multipath adaptive traffic engineering between an ingress node (source) and an egress node (destination). Its novel idea is to avoid network congestion and attacks that might exist in edge and node disjoint paths between two core network nodes.

This paper builds an adaptive, robust, and reliable traffic engineering scheme for better performance of communication network operations. This will also provision  quality of service (QoS) and protection of  traffic engineering to maximize network efficiency. Specifically,  we present a new approach, S-MATE  (secure MATE) is developed to  protect the network traffic between two core nodes (routers or switches) in a cloud network. S-MATE secures against a single link attack/failure by adding redundancy in one of the operational paths between the sender and receiver. The proposed scheme can be  built to secure core networks  such as optical and IP networks.
\end{abstract}

\section{Introduction}\label{sec:intro}

There have been several proposals to adapt the traffic between core network nodes  in Internet Service Provider (ISP) networks~\cite{kandula05,elwalid02,he06}.  Elwalid \emph{et al.}~\cite{elwalid02}  proposed an algorithm for multipath adaptive traffic engineering between an ingress node (source) and an egress node (destination) in a communication network. Their novel idea is to avoid network congestion that might exist in disjoint paths between two core  network nodes. They suggested load balancing among paths based on measurement and analysis of path congestion by using Multi-Protocol Label Switching (MPLS). MPLS  is an emerging tool for facilitating traffic engineering unlike explicit routing protocols that allow certain routing methodology from hop-to-hop in a network with multiple core devices. The major advantage of MATE is that it does not require scheduling, buffer management, or traffic priority in the nodes.

Network coding is a powerful tool that has been recently used to increase
the throughput, capacity, and performance of wired and wireless communication networks.
Information theoretic aspects of network coding have been investigated
in several research papers, see for example~\cite{ahlswede00,soljanin07,yeung06}, and the list of references therein. It offers benefits in terms of
energy efficiency, additional security, and reduced delay. Network coding
allows the intermediate nodes not only to forward packets using network
scheduling algorithms, but also to encode/decode them through algebraic
primitive operations~\cite{ahlswede00,fragouli06,soljanin07,yeung06}. For example, data loss
because of failures in communication links can be detected and recovered
if the sources are allowed to perform network coding operations~\cite{cai06,gkantsidis06,jaggi07}.

\begin{figure}[t]
\begin{center}
  \includegraphics[scale=0.65]{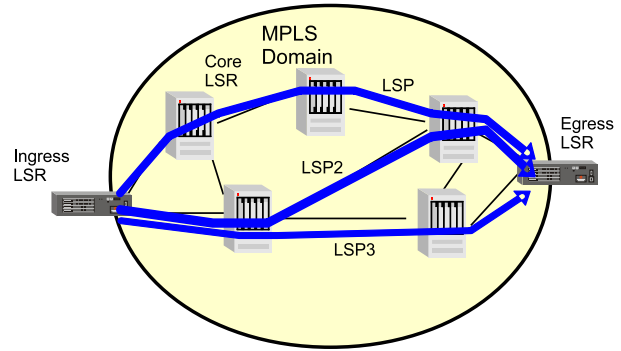}
  \caption{The network model is represented by two network nodes, an ingress node (source) and an egress node (receiver). There are $k$ link disjoint paths between the ingress and egress nodes.}
  \label{fig:netfig1}
  \end{center}
\end{figure}

MATE, which was previously proposed by one of the authors of this paper, is a traffic load balancing scheme that is suitable for S-MATE (secure MATE) as will be explained later.  MATE distributes traffic among the edge disjoint paths, so as to equalize the path delays. This is achieved by using adaptive algorithms. MATE inspired other traffic engineering solutions such as TexCP~\cite{kandula05} and the measurement-based optimal routing solution~\cite{guven06}.
In this paper, we will design a security scheme by using network coding to protect against an entity who cannot only copy/listen to the message, but  can also fabricate new messages or modify the current ones. We aim to build an adaptive, robust, reliable traffic engineering scheme for better performance and operation of communication networks. The scheme will also provide provisioning  quality of service (QoS) and protection of traffic engineering.

The rest of the paper is organized as follows. In Section~\ref{sec:model}, we present the network model and assumptions. In Sections~\ref{sec:MATE} and~\ref{sec:SMATE}, we review the MATE scheme and propose a secure MATE scheme based on network coding.  Section~\ref{sec:distributedcapacities} provides network protection using distributed capacities and QOS, and finally Section~\ref{sec:conclusion} concludes the paper.

\section{Network Model and Assumptions}\label{sec:model}

The network model can be represented as follows. Assume a given network represented by a set of nodes and links.  The network nodes are core nodes that transmit outgoing packets to the neighboring nodes in certain time slots.  The network nodes are ingress and egress nodes that share multiple edge and node disjoint paths.

We assume that the core nodes share $k$ edge disjoint paths, as shown in Fig.~\ref{fig:netfig1}, for one particular pair of ingress and egress nodes. Let $N=\{N_1,N_2,...\}$  be the set of nodes (ingress and egress) and let $L=\{L_{\ell h}^1,L_{\ell h}^2,...,L_{\ell h}^k\}$ be the set of paths from ingress node $N_{\ell}$ to an egress node $N_{h}$.  Every path $L_{\ell h}^i$ carries  segments of independent packets from  ingress node $N_\ell$ to egress node $N_h$. Let $P^{ij}_{\ell h}$ be the packet sent from the ingress node $N_\ell$ in path $i$ at time slot $j$ to the egress $N_h$. For  simplicity, we describe our scheme for one particular pair of ingress and egress nodes. Hence, we use $P^{ij}$ to represent a packet in path $i$ at time slot $j$.

Assume there are $\delta$ rounds (time slots) in a transmission session.  For the remaining paper, rounds and time slots will be used interchangeably . Packet $P^{ij}$ is indexed as follows:
\begin{eqnarray}\label{eq:plainpacket}
Packet_{\ell h}^{ij}(ID_{N_\ell}, X^{ij},round_j),
\end{eqnarray}
where $ID_{N_\ell}$ and $X^{ij}$ are, respectively, the sender ID and transmitted data from $N_\ell$ in the path $L_i$ at time slot $j$.
There are two types of packets: plain and encoded packets. A plain packet contains the unencoded data from the ingress to egress nodes as shown in Equation~(\ref{eq:plainpacket}). An encoded packet contains encoded data from different incoming packets. For example, if there are $k$ incoming packets to the ingress node $N_l$, then the encoded data traversed in the protection path $L_{l h}^i$ to the egress node $N_h$ are given by

\begin{eqnarray}\label{eq:encodeddata}
y^j=\sum_{i=1, j\neq i}^k P_{l h}^{ij},
\end{eqnarray}
where the summation denotes the binary addition. The corresponding packet becomes
\begin{eqnarray}\label{eq:encodedpacket}
Packet_{\ell h}^{ij}(ID_{N_\ell}, y^{j},round_j).
\end{eqnarray}

The following definition describes the \emph{working} and
\emph{protection} paths between two network switches as shown in
Fig.~\ref{fig:netfig1}.

\begin{definition}
The \emph{working paths} in a network with $k$ connection paths carry
un-encoded (plain) traffic under normal operations. The \emph{protection paths} provide
alternate backup paths to carry encoded traffic. A
protection scheme ensures that data sent from the sources will reach the
receivers in case of attack/failure incidences in the working paths.
\end{definition}

We make the following assumptions about the transmission of the plain and encoded packets.
\begin{compactenum}[i)]
\item The TCP protocol will handle the transmission and packet headers in the edge disjoint paths from  ingress to egress nodes.
\item The data from the ingress nodes  are sent in rounds and sessions throughout the edge disjoint paths to the egress nodes. Each session is quantified by the
    number of rounds (time slots) $n$. Hence, $t_j^\delta$ is the transmission time at the time
    slot $j$ in session $\delta$.

\item The attacks and failures on a path $L^i$ may be incurred by a network incident such as an eavesdropper, link replacement, and overhead. We
    assume that the receiver is able to detect a failure (attacked link), and our
    protection strategy described in S-MATE is able to recover it.
\item We assume that the ingress and egress nodes share a set of $k$ symmetric     keys. Furthermore, the plain and encoded data are encrypted by using this set of keys. That is $$x^i=Encypt_{key_i} (m^i),$$ where $m^i$ is the message encrypted by the $key_i$. Sharing symmetric keys between two entities (two core network nodes) can be achieved by using key establishment protocols described in~\cite{menezens01,schneier96}.
\item In this network model, we consider only a single link failure or attack; it is thus sufficient to apply the encoding and decoding operations over a finite
    field with two elements,  denoted as $\F_2=\{0,1\}$.
\end{compactenum}

 The traffic from the ingress node to the egress node in edge disjoint paths can be exposed to edge failures and network attacks. Hence, it is desirable to protect and secure this traffic. We assume that there is a set of $k$ connection paths that need to be fully guaranteed and protected
 against  a single  edge failure from ingress to egress nodes. We assume that all
connections have the same bandwidth, and each link (one hop or circuit) has the same bandwidth as the path.


\begin{figure}[t]
\begin{center}
  \includegraphics[scale=0.6]{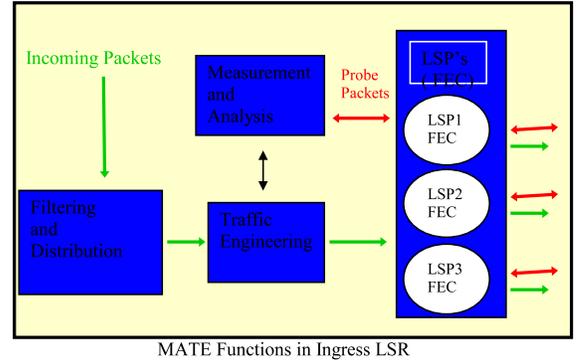}
  \caption{MATE traffic engineering at the ingress node.}
  \label{fig:netfig2}
  \end{center}
\end{figure}

\section{MATE Protocol}\label{sec:MATE}
MPLS (Multipath Protocol Label Switching) is an emerging tool for facilitating traffic engineering and out-of-band control. {  unlike explicit routing protocols, which allow certain routing methodology from hop-to-hop in a network with multiple core devices.}
As shown in Fig.~\ref{fig:netfig2}, MATE assumes that several explicit paths between an ingress node and an egress node in a cloud network  have been established. This is a typical setting which exists in operational Internet Service Provider (ISP) core networks (which implement MPLS).  The goal of the ingress node is to distribute traffic across the edge disjoint paths, so that the loads are balanced. One advantage of this load balancing is to equalize path delays, and to minimize traffic congestion~\cite{elwalid02,elwalid01}.

The following are the key features of the MATE algorithm.
\begin{compactenum}[1)]
\item The traffic is distributed at the granularity of the IP flow level. This ensures that packets from the same flow follow the same path, and hence there is no need for packet re-sequencing at the destination.   This is easily and effectively  achieved by using a hashing function on the five tuple IP address.
\item
MATE  is a traffic load balancing scheme, which is suitable for S-MATE,  as will be explained later.  MATE distributes traffic among the edge disjoint paths, so as to equalize the path delays. This is achieved by using adaptive algorithms as shown in Fig.~\ref{fig:netfig2} and Reference~\cite{elwalid02}

\item It is shown that  distributed load balancing (for each ingress egress pair) is stable and provably convergent. MATE assumes that several network nodes exist between ingress nodes as traffic senders, and egress nodes as traffic receivers.  Furthermore, the traffic can be adapted by using switching protocols such as CR-LDP~\cite{rosen98} and RSVP-TE\cite{bertsekas98}. An ingress node is responsible to manage the traffic in the multiple paths to the egress nodes so that  traffic congestion and overhead are minimized.
\end{compactenum}

As shown in Fig.~\ref{fig:netfig2}, Label Switch Paths (LSPs) from the ingress node to the egress node are provisioned before the actual packet transmissions occur.  Then, once the transmissions start, the ingress node will estimate the congestion that might occur in one or more of the $k$ edge disjoint paths. As stated in~\cite{elwalid02}, the congestion measure is related to one of the following factors: delay, loss rate, and bandwidth.  In general, each ingress node in the network will route the incoming packets into the $k$ disjoint paths. One of these paths will carry the encoded packets, and  all  other $k-1$ paths will carry plain packets. Each packet has its own routing number, so that the egress node will be able to manage the order of the incoming packet, and thus achieve the decoding operations.

MATE  works in two phases~\cite{elwalid02}: a monitoring phase and a load balancing phase.  These two phases will monitor the traffic and balance packets among all disjoint paths. One good feature of MATE is that its load balancing algorithms equalize the derivative of delay among all edge disjoint paths from an ingress node to an egress node. Furthermore, MATE's load balancing  preserves packet ordering since load balancing is done at the flow level (which is identified by a 5-tuple IP address) rather than at the packet level.
%

\section{S-MATE Scheme}\label{sec:SMATE}
In this section, we provide a scheme for securing MATE, called S-MATE (Secure Multipath Adaptive Traffic Engineering). The basic idea of S-MATE can be described  as shown in Equation~(\ref{eq:secMATE}). S-MATE inherits the traffic engineering components described in the previous section and in~\cite{elwalid02,elwalid01}.

Without loss of generality, assume that the network traffic between a pair of ingress and egress nodes is transmitted in $k$
edge disjoint paths, each of which carries  different packets. For simplicity, we assume that the number of edge disjoint paths and the number of rounds in one transmission session are equal.
\begin{figure}[t]
\begin{eqnarray}\label{eq:secMATE}
\begin{array}{|c|cccccc|c|}
\hline
& \multicolumn{6}{|c|}{\mbox{rounds from ingress to egress nodes}}&\ldots  \\
\hline
&1&2&3&\ldots&\ldots&n&\!\!\ldots\\
\hline    \hline
  L_{lh}^1& y^1&P^{11}&P^{12}&\ldots&\ldots&P^{1(n-1)} &\ldots   \\
    L_{lh}^2 &  P^{21}& y^2& P^{22}&\ldots&\ldots&\!\! P^{2(n-1)} &\ldots \\
L_{lh}^3 &  P^{31}& P^{32}&y^3& \ldots&\ldots&\!\! P^{3(n-1)} &\ldots \\
     \vdots&\vdots&\vdots&\vdots&\vdots&\vdots&\vdots&\ldots\\
     L_{lh}^j & P^{j1}&P^{j2}&\ldots&y^j&\ldots&\!\!P^{j(n-1)}&\ldots\\
 \vdots&\vdots&\vdots&\vdots&\vdots&\vdots&\vdots&\ldots\\
   L_{lh}^k & P^{k1}&P^{k2}&\ldots&\ldots&\!\!P^{k(k-1)}&y^n&\ldots\\
\hline
\hline
\end{array}
\end{eqnarray}
\end{figure}
There are  two types of packets:
\begin{compactenum}[i)]
\item {\bf Plain Packets:} These are packets $P^{ij}$ sent without coding, in which the ingress node does not need
    to perform any coding operations. For example, in the case of packets
    sent without coding, the ingress node  $N_l$ sends the following packet
    to the egress node  $N_h$:
\begin{eqnarray}
packet_{N_l \rightarrow N_h}(ID_{N_l},x^{ij},t_\delta^j),~for~ i=1,2,..,k, i\neq j.
\end{eqnarray}
The plain data $x^{ij}$ are actually the encryption of the message $m^{ij}$ obtained by using any secure symmetric encryption algorithm~\cite{menezens01}. That is, $x^{ij}=Encypt_{key_i} (m^{ij})$, where $key_i$ is a  symmetric key shared between $N_l$ and $N_h$.

\item  {\bf Encoded Packets:} These are packets $y^i$ sent with encoded data, in which the ingress node $N_l$
    sends other incoming data. In this case, the ingress node $N_l$ sends the following packet to egress node
$N_h$:
\begin{eqnarray}
packet_{N_l \rightarrow N_h}(ID_{N_l},
\sum_{i=1}^{j-1} x^{i~j-1}+\sum_{i=j+1}^k x^{ij},t^j_\delta).
\end{eqnarray}
The encoded packet will be used in case  any  of the working paths is compromised. The egress node will be able to detect the compromised data, and can recover them by using the data sent in the protection path.
\end{compactenum}

\begin{lemma}
The S-MATE scheme  is optimal against a single  link attack.
\end{lemma}

What we mean by optimal is that the encoding and decoding operations are achieved over the binary field with the least computational overhead. That is, one cannot find a better scheme than this proposed encoding scheme in terms of encoding operations. Indeed, one single protection path is used in case of a single attack path or failure.
The transmission is done in rounds (time slots), and hence  linear combinations of data
have to be from the same round time. This can be achieved by using the
time slot that is included in each packet sent by the ingress node.

\begin{lemma}
The network capacity between the ingress node and the egress node is given by $k-1$ in the case of one single attack path.
\end{lemma}

\noindent \textbf{Encoding Process:} There are several scenarios where the
encoding operations can be achieved. The encoding and decoding operations
will depend mainly on the network topology; i.e., on how the senders and receivers
are distributed in the network.
\begin{itemize}
\item The encoding operation is performed at only one ingress node $N_l$. In this case,
     $N_l$ will prepare and send the encoded data over $L_{lh}^i$ to the receiver $N_h$.
\item We assume that $k$ packets will be sent  in every transmission session from the ingress node. Also, if the number of incoming packets is greater than $k$, then a mod function is used to moderate the outgoing traffic in $k$ different packets. Every packet will be sent in a different path.
    \item Incoming packets with large sizes will be divided into small chunks, each with an equal size.
\end{itemize}

\noindent \textbf{Decoding Process:}
The decoding process is performed in a similar way as explained in the previous works~\cite{aly08preprint1,aly09-4}.

\begin{figure}[t]
\begin{center}
  \includegraphics[scale=0.70]{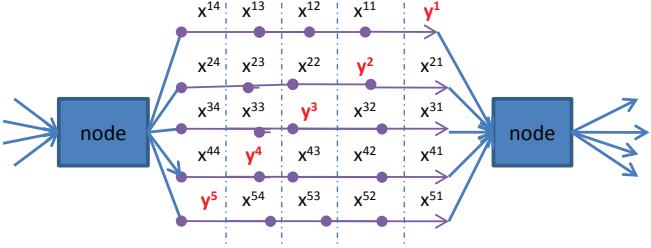} 
  \caption{Working and protection edge disjoint paths between two core nodes. The protection path carries encoded packets from all other working paths between ingress and egress nodes.}
  \label{fig:netfig3}
  \end{center}
\end{figure}

We assume that the ingress node $N_l$  assigns
the paths that will carry plain data { as shown in
Fig.~\ref{fig:netfig3}.} In addition, $N_l$ will encode the
data from all incoming traffics and send them in one path.
This will be used to protect  any  single link  attacks/failures. The objective is to withhold rerouting signals or transmitted packets due to link attacks. However, we provide strategies that use network coding and reduced capacity at the ingress nodes. We assume that the source nodes (ingress) are able to perform encoding operations and the receiver nodes (egress) are able to perform decoding operations.\\

One of S-MATE's objectives is to minimize the delay of the transmitted packets. So, the packets from one IP address will be received in order on one path. The key features of S-MATE can be described as follows:
\begin{itemize}
\item The traffic from the ingress node to the egress node is secured against eavesdroppers and intruders.

    \item No extra paths in addition to the existing network edge disjoint paths are needed to secure the network traffic.
        \item It can be implemented without adding new hardware or network components.

\end{itemize}

The following example illustrates the plain and encoded data transmitted from five senders to five receivers.

\medskip

\begin{example}
Let $N_l$ and $N_h$ be two core network nodes (sender and receiver) in a cloud network. Equation~(\ref{eq:example1}) explains the plain and encoded data sent in five consecutive time slots from the  sender to the  receiver. In the first time slot, the first connection carries encoded data, and all other connections carry plain data. Furthermore, the encoded data are distributed among  all connections in the time slots 2, 3, 4 and 5.
\begin{eqnarray}\label{eq:example1}
\begin{array}{|c|ccccc|c|c|}
\hline
cycle& &~~~~1&&&&2&3\\
\hline
rounds&1&2&3&4&5&\ldots&\ldots\\
\hline
\hline
L_{lh}^1 &y^1&x^{11}  &x^{12} &x^{13} &x^{14}& \ldots&\ldots\\
L_{lh}^2 &x^{21}&y^2  &x^{22} &x^{23}&x^{24}&\ldots&\ldots\\
L_{lh}^3 &x^{31}  &x^{32}&y^3 &x^{33}&x^{34}&\ldots&\ldots \\
L_{lh}^4    &x^{41} &x^{42} & x^{43}&y^{4}&x^{44}&\ldots&\ldots\\
L_{lh}^5    &x^{51} &x^{52} & x^{53}&x^{54}&y^5&\ldots&\ldots\\
\hline
\end{array}
\end{eqnarray}
The encoded data $y^j$, for $1 \leq j \leq 5$,  are sent as

\begin{eqnarray}
y^j=\sum_{i=1}^{j-1} x^{i~j-1}+\sum_{i=j+1}^5 x^{ij}.
\end{eqnarray}
We notice that every message has its own time slot. Hence, the protection data are distributed among all paths for fairness.
\end{example}


\section{A Strategy Against two attacked Paths}
In this section, we propose a strategy against two attacked paths (links), securing MATE against two-path attacks. The strategy is achieved using network coding and dedicated paths.  Assume we have $n$ connections carrying data from an ingress node to an egress node. All connections represent disjoint paths.

We will  provide two backup paths to secure against any two disjoint paths,
which might experience any sort of attacks. These two protection paths can be chosen
using network provisioning. The protection paths are fixed for all rounds
per session from the ingress node to the egress node, but they may vary among sessions. For example, the ingress node $N_l$  transmits a message $x^{i\ell}$ to  the egress node $N_h$ through path $L_{\ell h}^i$ at time $t_\delta^\ell$ in round time $\ell$ in session $\delta$. This process is explained
in Equation~(\ref{eq:n-1protection2}) as:

\begin{eqnarray}\label{eq:n-1protection2}
\begin{array}{|c|ccccc|c|}
\hline
& \multicolumn{5}{|c|}{\mbox{ cycle 1 }}&\ldots    \\
\hline
&1&2&3&\ldots&n&\!\!\ldots\\
\hline    \hline
 L_{lh}^1 & x^{11}&x^{12} &x^{13}&\ldots  &x^{1n} &\ldots \\
   L_{lh}^2 &  x^{21}&  x^{22}&x^{23}&\ldots&x^{2n} &\ldots \\
L_{lh}^3 &  x^{31}& x^{32}& x^{33}&\ldots&x^{3n} &\ldots \\
 \vdots&\vdots&\vdots&&\vdots&\vdots&\ldots\\
L_{lh}^i  & x^{i1} & x^{i2} &x^{i3}&\ldots& x^{in}&\ldots\\
     L_{lh}^j & y^{j1}&y^{j2}&y^{j3}&\ldots&y^{jn}&\ldots\\
    L_{lh}^k  & y^{k1}&y^{k2}&y^{k3}&\ldots&y^{kn}&\ldots\\
     L_{lh}^{i+1}& x^{(i+1)1} & x^{(i+1)2} &x^{(i+1)3}&\ldots& x^{(i+1)n}&\ldots\\
 \vdots&\vdots&\vdots&\vdots&\vdots&\vdots&\ldots\\
  L_{lh}^n & x^{n1}&x^{n2}&x^{n3}&\ldots&x^{nn}&\ldots\\
\hline
\hline
\end{array}
\end{eqnarray}
All $y_j^\ell$'s are defined as:
\begin{eqnarray} y^{j\ell}=\sum_{i=1,i\neq j \neq k}^n a_i^\ell x^{i\ell} \mbox{  and  } y^{k\ell}=\sum_{i=1,i\neq k \neq j}^n b_i^\ell x^{i\ell}.
\end{eqnarray}

The coefficients $a_i^\ell$ and $b_i^\ell$ are chosen over a finite field
$\F_q$ with $q > n-2$, see~\cite{aly08preprint1} for more details. One way to choose these coefficients is by using the follow two vectors.
\begin{eqnarray}\label{eq:twovectors}
\left[\begin{array}{ccccc}
1&1&1&\ldots&1\\
1&\alpha&\alpha^2&\ldots&\alpha^{n-3}
\end{array}\right]
\end{eqnarray}
Therefore, the coded data is
\begin{eqnarray} y^{j\ell}=\sum_{i=1,i\neq j \neq k}^n  x^{i\ell} \mbox{  and  } y^{k\ell}=\sum_{i=1,i\neq k \neq j}^n \alpha^{i \mod n-2} x^{i\ell}.
\end{eqnarray}
In the case of two failures, the receivers will be able to solve two linearly independent equations in two unknown variables. For instance, assume the two failures occur in paths number two and four. Then the receivers will be able to construct two equations with coefficients
\begin{eqnarray}\label{eq:twovectors2}
\left[\begin{array}{cc}
1&1\\
\alpha&\alpha^{3}
\end{array}\right]
\end{eqnarray}
 Therefore, we have
\begin{eqnarray}
x^{2\ell}+x^{4\ell}\\
\alpha x^{2\ell}+\alpha^3 x^{4\ell}
\end{eqnarray}
One can multiply the first equation by $\alpha$ and subtract the two equations to obtain value of $x^{4\ell}$.

We  notice that the encoded data symbols
$y^{j\ell}$ and $y^{k\ell}$ are fixed per one session but it is
varied for other sessions. This means that the path $L_{lh}^j$ is dedicated to
send all encoded data $y^{j1},y^{j2},\ldots,y^{jn}$.
\begin{lemma}
The  network capacity of the protection strategy against two-path attacks is given by
$n-2$.
\end{lemma}

There are three different scenarios for two-path attacks, which
can be described as follows:
\begin{compactenum}[i)]
\item If the two-path attacks occur in the backup protection paths $L_{lh}^j$
    and
    $L_{lh}^k$,
    then no recovery operations are required at the egress node.
\item If the two-path attacks  occur in one backup protection path say
    $L_{lh}^j$ and one working path $L_{lh}^i$, then  recovery operations are
    required.
\item If the two-path attacks occur in two working paths, then in this
    case the two protection paths are used to recover the lost data. The
    idea of recovery in this case is to build a system of two linearly independent equations
    with two unknown variables.
\end{compactenum}

\section{Network Protection Using Distributed Capacities and QoS}\label{sec:distributedcapacities}

In this section, we develop a network protection strategy in which  some connection paths have high priorities (less bandwidth and high demand). Let $k$ be  the set of available connections (disjoint paths from ingress to egress nodes). Let $m$ be the set of rounds in every  cycle. We assume that all connection paths might not have the same priority demands and working capacities.   Connections that carry applications with multimedia traffic have higher priorities than those of applications carrying data traffic.. Therefore, it is required to design network protection strategies based on the traffic and sender priorities.

Consider that available working connections $k$ may use their bandwidth assignments in asymmetric ways. Some connections are less demanding in terms of bandwidth requirements than other connections that require full capacity frequently. Therefore, connections with less demand can transmit more protection packets, while other connections demand more bandwidth, and can therefore transmit fewer protection packets throughout the transmission rounds. Let $m$ be the number of rounds and $t_i^\delta$ be the time of transmission in a cycle $\delta$ at round $i$. For a particular cycle $i$, let $t$ be the number of protection paths against $t$ link failures or  attacks that might affect the working paths. We will design a network protection strategy against $t$ arbitrary link failures  as follows. Let the source $s_j$ send $d_i$ data packets and $p_i$ protection packets such that $d_j+p_j=m$. Put differently:

\begin{eqnarray}
\sum_{i=1}^k (d_i+p_i)=km.
\end{eqnarray}
In general, we do not assume that $d_i =d_j$ and $p_i=p_j$.

\begin{figure}
\begin{eqnarray}\label{eq:tfailures2}
\begin{array}{|c|ccccccc|}
\hline
& \multicolumn{7}{|c|}{\mbox{ round time cycle 1 }} \\ \hline
\hline
&1&2&3&4&\ldots&m-1&m\\
\hline    \hline
L_{\ell h}^1 & y^{11}&x^{11} &x^{12}&y^{12}&\ldots &y^{1p_1}&x^{1d_1} \\
L_{\ell h}^2 &  x^{21}& y^{21} &x^{22}&x^{23}&\ldots&x^{2d_2} &y^{2p_2} \\
\vdots&\vdots&\vdots&\vdots&\vdots&\vdots&\vdots&\vdots\\
L_{\ell h}^{i} &  y^{i1}&  x^{i1} &x^{i2}&y^{i2}&\ldots&y^{ip_i}  &x^{id_i}\\
\vdots&\vdots&\vdots&\vdots&\vdots&\vdots&\vdots&\vdots\\
L_{\ell h}^{j} &  x^{j1}&  x^{j2} &y^{j1}&x^{j3}&\ldots&x^{jd_j}&y^{jp_j}  \\
\vdots&\vdots&\vdots&\vdots&\vdots&\vdots&\vdots&\vdots\\
L_{\ell h}^k & x^{k1}&y^{k1}&x^{k2}&x^{k4}&\ldots&y^{kp_k}&x^{kd_k}\\
\hline
\end{array}
\end{eqnarray}
\end{figure}
The encoded data $y^{i\ell}$ are given by
\begin{eqnarray}
y^{i\ell} =\sum_{k=1,y^{k\ell} \neq y^{k\ell}} x^{k\ell}.
\end{eqnarray}

\begin{figure}[t]
\begin{center}
  \includegraphics[width=8.5cm,height=4cm]{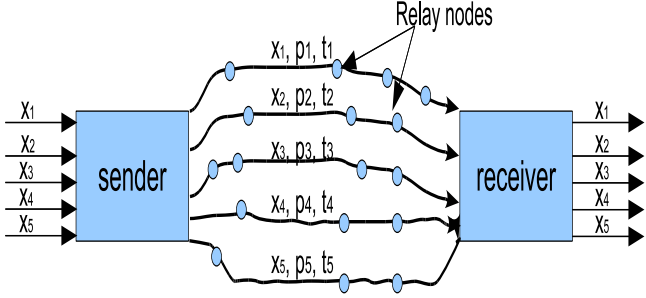}
  \caption{Working and protection edge disjoint paths between two core nodes (ingress and egress nodes). Every path $L_i$ carries encoded and plain packets depending on the traffic priority $p_i$ and time $t_i$.}
  \label{fig:netfig4}
  \end{center}
\end{figure}

We assume that the maximum number of attacks/failures that might occur in a particular cycle is $t$. Hence, the number of protection paths (paths that carry encoded data) is $t$. The selection of the working and protection paths in every round is performed by using a demand-based priority function at the senders's side. It will also depend on the traffic type and service provided on these protection and working connections. See Fig.~\ref{fig:netfig4} for ingress and egress nodes with five disjoint connections.

In Equation~(\ref{eq:tfailures2}), every connection $i$ is used to carry $d_i$ unencoded data $x^{i1},x^{i2},\ldots,x^{id_i}$ (working paths) and $p_i$ encoded data $y^{i1},y^{i2},\ldots,y^{ip_i}$ (protection paths) such that $d_i+p_i=m$.

\begin{lemma}\label{lem:nps-t2}
Let $t$ be the number of connection paths carrying encoded data in every round. Then, the  network capacity $C_\N$ is given by
\begin{eqnarray}
C_\N=k-t.
\end{eqnarray}
\end{lemma}
\begin{IEEEproof}
The proof is straightforward from the fact that $t$ protection paths exist in every round, and hence $k-t$ working paths are available throughout all $m$ rounds.
\end{IEEEproof}

\medskip

\section{Conclusion}\label{sec:conclusion}
In this paper, we have proposed the S-MATE scheme (secure multipath adaptive traffic engineering) for operational networks. We have used network coding of transmitted packets to protect the traffic between two  core nodes (routers, switches, etc.) that could exist in a cloud network. Our assumption is based on the fact that core network nodes share multiple edge disjoint paths from the sender to the receiver. S-MATE  can secure network traffic against  single link attacks/failures by adding redundancy in one of the operational paths. Furthermore, the proposed scheme can be  built to secure operational networks  including optical and multipath adaptive networks. In addition, it can provide security services at the IP and data link layers.

\scriptsize
\bibliographystyle{plain}

\bibliographystyle{ieeetr}

\begin{thebibliography}{20}

\bibitem{ahlswede00}
R.~Ahlswede, N.~Cai, S.-Y.~R. Li, and R.~W. Yeung.
\newblock Network information flow.
\newblock {\em IEEE Trans. Inform. Theory}, 46(4):1204--1216, 2000.

\bibitem{aly09-4}
S.~A. Aly and A.~E. Kamal.
\newblock Network coding-based protection strategy against node failures.
\newblock {\em Proc. IEEE Int'l Conf. Commun. (ICC'09) confernce}, Dresden,
  Germany, June 2009.

\bibitem{aly08preprint1}
S.~A. Aly and A.~E. Kamal.
\newblock Network protection codes: Providing self-healing in autonomic
  networks using network coding.
\newblock {\em IEEE Transaction on Networking}, submitted, 2009.
\newblock arXiv:0812.0972v1 [cs.NI].

\bibitem{bertsekas98}
D.~P. Bertsekas and J.~N. Tsitsiklis.
\newblock {\em Parallel and Distributed Computation}.
\newblock Prentice-Hall, 1989.

\bibitem{cai06}
N.~Cai and W.~Yeung.
\newblock Network error correction, part 2: Lower bounds.
\newblock {\em Communications in Information and Systems}, 6:37--54, 2006.

\bibitem{rosen98}
A.~Viswanathan E.~C.~Rosen and R.~Callon.
\newblock Multiprotocol label switching architecture.
\newblock Internet draft $<$draft-ietf-mpls-arch-01.txt$>$, March 1998.

\bibitem{elwalid02}
A.~Elwalid, C.~Jin, S.~Low, and I.~Widjaja.
\newblock Mate: Multipath adaptive traffic engineering.
\newblock {\em The International Journal of Computer and Telecommunications
  Networking}, 40(6):695--709, 2002.

\bibitem{elwalid01}
A.~Elwalid, C.~Jin, S.~Low, and I.~Widjaja.
\newblock {MATE}: {MPLS} adaptive traffic engineering.
\newblock In {\em Proc. IEEE {INFOCOMM}}, Anchorage, Alaska, April 22-26, 2001.

\bibitem{fragouli06}
C.~Fragouli, J.~Le Boudec, and J.~Widmer.
\newblock Network coding: An instant primer.
\newblock {\em ACM SIGCOMM Computer Communication Review}, 36(1):63--68, 2006.

\bibitem{soljanin07}
C.~Fragouli and E.~Soljanin.
\newblock {Network Coding Applications, Foundations and Trends in Networking}.
\newblock {\em Hanover, MA, Publishers Inc.}, 2(2):135--269, 2007.

\bibitem{gkantsidis06}
C.~Gkantsidis and P.~Rodriguez.
\newblock Cooperative security for network coding file distribution.
\newblock In {\em Proc. IEEE INFOCOM}, Barcelona, Catalunya, Spain, April,
  2006.

\bibitem{he06}
J.~He, M.~Chiang, and J.~Rexford.
\newblock {DATE: Distributed Adaptive Traffic Engineering}.
\newblock In {\em Proc. IEEE INFOCOMM}, volume~3, Barcelona, Catalunya, Spain,
  April, 2006.

\bibitem{jaggi07}
S.~Jaggi, M.~Langberg, S.~Katti, T.~Ho, D.~Katabi, and M.~Medard.
\newblock Resilient network coding in the presence of byzantine adversaries.
\newblock In {\em Proc. IEEE INFOCOM}, Anchorage , Alaska , USA, April, 2007.

\bibitem{kandula05}
S.~Kandula, D.~Katabi, B.~Davie, and A.~Charny.
\newblock Walking the tightrope: Responsive yet stable traffic engineering.
\newblock In {\em ACM SIGCOMM}, Philadelphia, PA, August, 2005.

\bibitem{menezens01}
A.~J. Menezes, P.~C.~van Oorschot, and S.~A. Vanstone.
\newblock {\em Handbook of Applied Cryptography}.
\newblock CRC Press, Boca Raton, Fla., 2001.

\bibitem{schneier96}
B.~Schneier.
\newblock {\em Applied Cryptography}.
\newblock 2nd edition, Wiley, New York, 1996.

\bibitem{guven06}
M.A.~Shayman T.~Guven, R.J.~La and B.~Bhattacharjee.
\newblock Measurement-based optimal routing on overlay architectures for
  unicast sessions.
\newblock {\em Computer Networks}, 50(12):1938--–1951, August 2006.

\bibitem{yeung06}
R.~W. Yeung, S.-Y.~R. Li, N.~Cai, and Z.~Zhang.
\newblock {\em Network Coding Theory}.
\newblock Now Publishers Inc., Dordrecth, The Netherlands, 2006.

\end{thebibliography}
\end{document}